\documentstyle[aps,twocolumn]{revtex}

\begin{document}
\author{I.M. Sokolov}
\author{Theoretische Polymerphysik, Universit\"{a}t Freiburg, Herman-Herder-Str.3,
D-79104 Freiburg im Breisgau, Germany}
\title{Thermodynamics and Fractional Fokker-Planck Equations}
\date{\today}
\maketitle

\begin{abstract}
The relaxation to equilibrium in many systems which show strange kinetics is
described by fractional Fokker-Planck equations (FFPEs). These can be
considered as phenomenological equations of linear nonequilibrium theory. We
show that the FFPEs describe the system whose noise in equilibrium funfills
the Nyquist theorem. Moreover, we show that for subdiffusive dynamics the
solutions of the corresponding FFPEs are probability densities for all cases
where the solutions of normal Fokker-Planck equation (with the same
Fokker-Planck operator and with the same initial and boundary conditions)
exist. The solutions of the FFPEs for superdiffusive dynamics are not always
probability densities. This fact means only that the corresponding kinetic
coefficients are incompatible with each other and with the initial
conditions.
\end{abstract}


\section{Introduction}

Different physical systems, like polymers chains, membranes, networks and
other generalized Gaussian structures, often show a long temporal memory due
to the complex hierarchical organization of the modes of their motion. On
the other hand, the response of these systems to an external perturbation
stays linear for a wide range of parameters \cite
{Douglas,Schiessel,Schies2,BKB}. As recently suggested, the response
dynamics is well-described by dynamical equations introducing fractional
time-derivatives instead of whole-number ones \cite
{Douglas,Schiessel,Schies2}. From the thermodynamical point of view, the
systems do not show any peculiarities close to equilibrium in contact with a
classical heat bath.

The relaxation to equilibrium in such systems is thus described by
fractional Fokker-Planck equations (FFPEs), which follow as phenomenological
linear response equations. The corresponding equations are especially
popular in application to a slow (subdiffusive) dynamics \cite{MeK} and
where introduced ad hoc much before the microscopic basis for such equations
got clear.

We show that the typical FFPEs with the fractional derivative in front of
the normal Fokker-Planck operator, 
\begin{eqnarray}
\frac{\partial }{\partial t}P(x,t) &=&~_{t_{0}}D_{t}^{1-\gamma }\mu \left[ 
\frac{\partial }{\partial x}f(x,t)P(x,t)\right.   \label{FFPE} \\
&&+\left. k_{B}T\frac{\partial ^{2}}{\partial x^{2}}P(x,t)\right] , 
\nonumber
\end{eqnarray}
are the only possible variant for description of nearly equilibrium systems
showing linear response, since they (and only they) fulfill the Nyquist
theorem which connects linear response behavior with the noise spectrum at
equilibrium. Here the fractional derivative operator $D_{t}^{1-\gamma }$ is
defined by 
\begin{equation}
_{t_{0}}D_{t}^{1-\gamma }W=\frac{1}{\Gamma (\alpha )}\frac{\partial }{%
\partial t}\int_{t_{0}}^{t}\frac{dt^{\prime }W(x,t^{\prime })}{(t-t^{\prime
})^{1-\gamma }}.  \label{FFPE1}
\end{equation}
The value of $\gamma =1$ corresponds to an identity transformation, leading
to the case of pure diffusive behavior, the cases with $\gamma <1$
correspond to subdiffusive behavior, and the case $\gamma >1$ to a
superdiffusive dynamics, like one considered in Ref.\cite{MeK1}.

Other forms of FFPEs are known, e.g. a Galilean-invariant form \cite{MeK,MKS}
which appears quite naturally when describing transport in a given velocity
field, and forms with different fractional time-derivatives in front of the
first and second spatial derivatives, which may appear as dynamical
equations in many other contexts (economics, biology, etc.). They do not
apply to cases of thermodynamical relaxation close to equilibrium.

We show that FFPEs which describe subdiffusive dynamics always have
thermodynamically sound solutions when the corresponding normal
Fokker-Planck equation also has them. Such solutions are subordinated to the
solution of a normal Fokker-Planck equation with the same initial/boundary
conditions. The situation with the superdiffusive dynamics is different:
here not all combinations of external potential, diffusion coefficient and
memory kernel give rise to physical solutions (positive probability
densities), as it is e.g. the case for a fractional generalization of
diffusion with drift. We discuss why it is so and exemplify this situation
by processes subordinated to the solutions of a generic transport equation
(related to a Liouville equation).

\section{FFPE's as a phenomenological linear response theory}

Let us first discuss the properties of FFPEs as phenomenological equations
being very similar to normal, diffusive Fokker-Planck equation (FPE). Within
standard phenomenological linear nonequilibrium theory \cite{GroM,Balescu}
the diffusion equation in a weak external field (i.e. a forward
Fokker-Planck equation) follows as a consequence of local conservation of
probability, 
\begin{equation}
\frac{\partial P}{\partial t}=-div\ {\bf j}  \label{Concervation}
\end{equation}
and a phenomenological linear response assumption 
\begin{equation}
{\bf j}=\lambda ^{(1)}{\bf f}P-\lambda ^{(2)}grad\ P,  \label{Fick}
\end{equation}
where $\lambda ^{(1)}=\mu $ and $\lambda ^{(2)}=D$ are the kinetic
coefficients (the mobility, and the diffusion coefficient, respectively).
The phenomenological interpretation of the second equation is that the
current in our system can be caused by weak external field (and follows the
Ohm's law) and by concentration gradient (the first Fick's law), and that
both effects are independent as long as deviations from equilibrium are
small.

In general, the linear response can be retarded and then follows the
equation 
\begin{equation}
{\bf j}(t)=\Phi _{t}^{(1)}\left\{ {\bf f}(t^{\prime })P(t^{\prime })\right\}
-\Phi _{t}^{(2)}\left\{ grad\ P(t^{\prime })\right\} .  \label{LinResp}
\end{equation}
Here $\Phi _{t}$ are typically causal integrals of convolution type: 
\begin{equation}
\Phi _{t}^{(i)}\left\{ f(t)\right\} =\int_{t_{0}}^{t}\varphi
^{(i)}(t-t^{\prime })f(t^{\prime })dt^{\prime },  \label{Convol}
\end{equation}
where the lower integration limit $t_{0}$ can be either finite or infinite.
Here we again assume behaviors typical for the systems close to equilibrium.
Inserting Eq.(\ref{LinResp}) into Eq.(\ref{Concervation}) we get a
nonmarkovian (nonlocal in time) Fokker-Planck equation of the from 
\begin{eqnarray}
\frac{\partial }{\partial t}P(x,t) &=&\hat{\Phi}_{t}^{(1)}\left[ -\frac{%
\partial }{\partial x}f(x,t)P(x,t)\right]   \label{FIS} \\
&&+\hat{\Phi}_{t}^{(2)}\left[ \frac{\partial ^{2}}{\partial x^{2}}%
P(x,t)\right]   \nonumber
\end{eqnarray}
(here we restrict ourselves to a one-dimensional case).

Evaluating the first moment $M_{1}(t)$ of the distribution $P(x,t)$ under
influence of a homogeneous force $f$ we get that the evolution of the
response follows the equation 
\begin{equation}
\overline{v}=\frac{\partial }{\partial t}M_{1}=\hat{\Phi}_{t}^{(1)}f
\label{LinR}
\end{equation}
from which it is clear that the operator $\hat{\Phi}_{t}^{(1)}$ is exactly
the one describing the linear response of the system, so that the inverse
operator corresponds to the system's impedance.

Let us consider the noise produced by our system at equilibrium. The fact
that the system is equilibrated means that it was created long ago, so that $%
t_{0}\rightarrow -\infty $. Let us consider a Green's function of the
equilibrium system, fulfilling the equation: 
\begin{eqnarray}
\frac{\partial }{\partial t}G(x,t) &=&-\hat{\Phi}_{t}^{(1)}\left[ \frac{%
\partial }{\partial x}fG(x,t)\right] +  \label{Green} \\
&&+\hat{\Phi}_{t}^{(2)}\left[ \frac{\partial ^{2}}{\partial x^{2}}%
G(x,t)\right] +\delta (x)\delta (t).  \nonumber
\end{eqnarray}
The Fourier-transform of the Green's function in both spatial and temporal
domain is given by: 
\begin{equation}
i\omega G=\left[ \Phi ^{(1)}(\omega )ikf+\Phi ^{(2)}(\omega )k^{2}\right]
G+1,
\end{equation}
having a solution 
\begin{equation}
G(k,\omega )=\frac{1}{i\omega +\Phi ^{(1)}(\omega )ikf+\Phi ^{(2)}(\omega
)k^{2}}.
\end{equation}
Now, we are interested in the power spectrum of the equilibrium ($f=0$)
noise generated by our system. Let us consider the second moment of $G$ in
frequency domain, $M_{2}(\omega )=-\frac{\partial ^{2}}{\partial k^{2}}%
\left. G(k,\omega )\right| _{k=0}=2\Phi ^{(2)}(\omega )/\omega ^{2}$. Note
that $x$ is the time-integral of the instantaneous velocity, so that the
power spectrum of velocity (current) is exactly 
\begin{equation}
S_{v}(\omega )=2\mbox{Re}\Phi ^{(2)}(\omega ).
\end{equation}
Note also that the noise at equilibrium fulfils the Nyquist theorem \cite
{Stratonovich}, according to which 
\begin{equation}
S_{v}(\omega )=2k_{B}T\mbox{Re}\Phi ^{(1)}(\omega ),
\end{equation}
so that the operators $\Phi ^{(1)}$ and $\Phi ^{(2)}$ are not independent: 
\begin{equation}
\mbox{Re}\Phi ^{(1)}(\omega )=k_{B}T\mbox{Re}\Phi ^{(2)}(\omega ),
\label{Nyk}
\end{equation}
which for $\Phi $-operators of the fractional derivative type imply that $%
\Phi ^{(2)}=k_{B}T\Phi ^{(1)}$. All equations with $\Phi ^{(2)}=k_{B}T\Phi
^{(1)}$ are thermodynamically sound: they fulfill the generalized Einstein
relation and describe the relaxation to a Boltzmann distribution, which
properties follow also from the microscopic description of the corresponding
generalized Gaussian structures \cite{BKB}. The equations with independent $%
\Phi ^{(1)}$ and $\Phi ^{(2)}$ will typically lead to behavior at variance
with predictions of equilibrium thermodynamics. 

Note that the most systems for which the fractional dynamics was applied are
''normal'' although complex situations like polymers, membranes or fractal
webs. In what follows we discuss only the case which describes such systems
close to thermal equilibrium, for which the generalizations of FPE like Eq.(%
\ref{FFPE}) can be considered as thermodynamically sound phenomenological
laws. We also note that equations like Eq.(\ref{FFPE}) can be derived within
the framework of stochastic approach \cite{Metz2000}, where they apply to
situations close to thermal equilibrium. On the other hand, the equations
with {\it different} temporal operators are also widely used: an example is
a Galilean invariant FFPE of Ref.\cite{MKS}. This equation appears quite
naturally when describing transport in a given velocity field, i.e. when our
system is in a contact with a strongly nonequilibrium flow of fluid (a river
instead of a bath!). Other variants with different orders of fractional
temporal derivatives may appear as dynamical equations in many other
contexts (economics, biology, etc.) but would never apply to the case of
thermodynamical relaxation in a system close to equilibrium, since they
violate Eq.(\ref{Nyk}). The situation with the systems whose dynamics shows
linear response but is described by the FFPEs of a type different from one
considered above is similar to one which arises when negative temperatures
are considered \cite{Abragam}: the systems described by such dynamics can
live as isolated systems but can not be in equilibrium with any ''normal''
macroscopic bath. Interacting with a heat bath, such systems will gain or
lose energy until they leave the linear response regime and get a noise
spectrum conformal with equilibrium (and with a Boltzmann energy
distribution). 

Note that the FFPEs like Eq.(\ref{FFPE}) were proposed for processes with
finite increments (like continuous time random walk processes) or ones with
continuous trajectories (fractional Brownian motion), situations for which
the assumption of the local (differential) conservation law is proved. The
related thermodynamical considerations show that a system whose noise does
not posses any second moment (L\'{e}vy-noise), does not fulfill local
conservation, Eq.(\ref{Concervation}), and can hardly exhibit linear
response, a fact found in Ref.\cite{Sok1} on an example of subordinated
processes. This case is addressed in Ref.\cite{SKB} and leads to a different
form of FFPEs with fractional {\it spatial} operators.

Considerations based on linear nonequilibrium thermodynamics are somewhat
too general, since Eqs.(\ref{Concervation}) and (\ref{LinResp}) guarantee
the overall conservation of the value $P$ but not the fact that this $P$ is
a nonnegative quantity. The same equations will apply for electric charge
and current (which can be of both signs and may oscillate) and for density
or temperature, which are essentially nonnegative. Thus, in order to check
that the corresponding equation is thermodynamically sound one has to prove
that if the initial condition corresponds to a nonnegative density $P(x,0)$,
the density $P(x,t)$ will stay nonnegative during all the following
evolution. Since we concentrate here on the properties of relaxation to
equilibrium, the force term and the diffusion coefficient in our system will
be considered time-independent.

The proof of the non-negativity of solution for the force-free case was
given in \cite{Schneider} for the subdiffusive case. We show that the same
is the case for the arbitrary external force. Namely, we shall show that all
solutions of FFPEs with $\gamma \leq 1$ in arbitrary time-independent
potential force field are thermodynamically sound, and describe the
transport of a positively defined density. Moreover, we show that
superdiffusive equations with $1<\gamma \leq 2$ do not always possesses
physically sound solutions, unless some additional conditions are fulfilled.
Fokker-Planck equations of the type of Eq.(\ref{FFPE}) with $\gamma >2$ seem
to contradict physical sense. However, the superballistic behavior (say,
L\'{e}vy flights) can be described by the FFPEs of a different class, see
Ref.\cite{SKB}.

\section{The subdiffusive case: temporal subordination}

Let us first consider the subdiffusion case, $0<\gamma <1$. Note that the
solution of {\it subdiffusive} FFPE under time-independent force can be put
in the following form: 
\begin{equation}
P(x,t)=\int_{0}^{\infty }F(x,\tau )T(\tau ,t)d\tau ,  \label{Subord}
\end{equation}
where 
\begin{equation}
T(\tau ,t)=\frac{t}{\gamma \tau ^{1+1/\gamma }}{\cal L}(t/\tau ^{1/\gamma
},\gamma ,-\gamma )  \label{Levy}
\end{equation}
with ${\cal L}(t/\tau ^{1/\gamma },\gamma ,-\gamma )$ being an extreme
(one-sided) L\'{e}vy-stable law of index $\gamma $ \cite{MeK,BaS}, and $%
F(x,\tau )$ is the solution of ''normal'' FPE under the same force and the
same initial conditions: 
\begin{equation}
\frac{\partial }{\partial t}F(x,t)=-\mu \frac{\partial }{\partial x}%
f(x)F(x,t)+k_{B}T\frac{\partial ^{2}}{\partial x^{2}}F(x,t).
\end{equation}
To check this let us take the Laplace-transform of both sides of Eq.(\ref
{FFPE}), and note that this transform acts only on the $t$ variable, which
appears in the Eq.(\ref{Subord}) as a parameter: The Laplace transform in $t$
of Eq.(\ref{Subord}) reads: 
\begin{eqnarray}
P(x,u) &=&\int_{0}^{\infty }F(x,\tau )u^{\gamma -1}\exp (-\tau u^{\gamma
})d\tau   \label{Pofu} \\
&=&u^{\gamma -1}\tilde{F}(x,u^{\gamma }).  \nonumber
\end{eqnarray}
The fractional temporal differentiation leads to 
\begin{eqnarray}
u^{\gamma }\tilde{F}(x,u^{\gamma })-P(x,+0) &=&-\mu \frac{\partial }{%
\partial x}f(x)\tilde{F}(x,u^{\gamma })+  \label{Lapla} \\
&&+\mu k_{B}T\frac{\partial ^{2}}{\partial x^{2}}\tilde{F}(x,u^{\gamma }). 
\nonumber
\end{eqnarray}
Note that for $t\rightarrow 0$ the $T$-functions are strongly concentrated,
so that $T(\tau ,t)\rightarrow \delta (\tau )$ and $P(x,+0)=F(x,+0)$.
Changing now to a new variable $\lambda =u^{\gamma }$ we recognize in Eq.(%
\ref{Lapla}) the Laplace-transform of the ''normal'' FPE with the same
time-independent force and the same initial conditions. Thus, the solution
of FFPE can be obtained from the solution of FPE by immediate integration.
Moreover, each functional of such solution (e.g. any moment) can be
immediately obtained by weighing the corresponding functional of the FPE
solution with a probability distribution, Eq.(\ref{Levy}) for which useful
analytic representations are known. Thus, the equations with $\gamma \leq 1$
in any (temporally constant) force field $f$ obey regular Boltzmann
thermodynamics and correspond to the transport of a positively defined
density. Our result generalizes the mathematical treatment of Schneider and
Wyss and shows that the solution of a FFPE describing subdiffusive transport
in external potential is a probability density whenever the solution of a
normal FPE in the same potential is one. The generalization to higher
dimensions is evident. Note that our discussion here parallels that of Ref.%
\cite{BaS} where the fractional Kramers equation is considered.

Note that Eq.(\ref{Lapla}) shows an extremely interesting property of free
relaxation of the systems described by subdiffusive FFPEs, namely the fact
that the solution of Eq.(\ref{FFPE}) having a form of convolution (linear
response with a long-time memory kernel) can be represented in the form of
subordination, i.e. they correspond to the behavior of the system whose
development is governed by its own internal clock, which is not synchronized
with our physical time \cite{Hilfer}. The first reasonable use of this fact
can be probably attributed to P.K. Clark, see Ref.\cite{Clark} for the
discussion of the role of subordination in economical processes. This
(operational) time is a variable which is Laplace conjugated to $u^{\gamma }$%
, and can be considered as a real time variable, since it is monotonously
growing in our physical time and allows to order the events sequentially.

\section{Some properties of temporal subordination}

The integral transform, Eq.(\ref{Subord}) will be called a subordination
transformation (ST), the term ''time-expanding transformation'' (TET) will
be reserved for those with $\gamma <1$. In order to sharpen the instruments
needed for understanding the consequences of Eq.(\ref{FFPE}) let us discuss
some properties of STs with $T(\tau ,t)$-functions from the class discussed
above. The physical time $t$ will be called the outer variable of the
function $T(\tau ,t)$, and the variable $\tau $ (operational time) over
which the integration is performed will be called its inner variable. Note
that the STs are just a type of transformation typically arising in a
context of separation of variables (the eigenfunctions decomposition with
integration or summation over the eigenfunctions numbered by the eigenvalue $%
\tau $). In the eigenfunction decomposition we start from the solution of a
time-independent equation (zero order in time) and rise the order of
temporal derivative (say to one) by applying an equation of the type of Eq.(%
\ref{Subord}). In the case of subdiffusive FFPE we proceed in the opposite
direction: we lower the order of a temporal derivative by applying
subordination. As we proceed to show, a superposition of two functions of
the type of $T(\tau ,t)$ (with indices $\gamma _{1}$ and $\gamma _{2}$)
discussed above, is again a function of the same class with the index $%
\gamma =\gamma _{1}\gamma _{2}$. Let us suppose that both indices, $\gamma
_{1}$ and $\gamma _{2}$ are less than unity. We know that the
Laplace-transform of $T(\tau ,t)$ in its outer variable reads: $T(\tau
,u)=u^{\gamma -1}\exp (-\tau u^{\gamma })$. We thus get: 
\begin{equation}
T^{*}(\tau ,u)=\frac{1}{\gamma _{1}}\int \frac{t^{\prime }}{\tau
^{1+1/\gamma _{1}}}{\cal L}_{\gamma _{1}}\left( \frac{t^{\prime }}{\tau
^{1/\gamma _{1}}}\right) u^{\gamma _{2}-1}\exp (-t^{\prime }u^{\gamma
_{2}})dt^{\prime },
\end{equation}
which is again a Laplace-transform of a $T$-function in its outer variable.
Using this fact once again we get 
\begin{eqnarray}
T^{*}(\tau ,u) &=&u^{\gamma _{2}-1}\int_{0}^{\infty }T_{1}(\tau ,\xi )\exp
(-\xi u^{\gamma _{2}})d\xi   \nonumber \\
&=&u^{\gamma _{2}}{}^{\gamma _{1}-1}\exp (-\tau u^{\gamma _{1}\gamma _{2}}).
\end{eqnarray}
Thus, parallel to the L\'{e}vy-case of Ref.\cite{Sok1}, the superposition of
two TETs is a TET again. Note that all $T$-functions with $\gamma <1$ are
probability densities in their inner variable: they are nonnegative and
integrable. On the other hand, the $T$-functions rising the order of the
temporal variable have a Laplace-transform in the outer variable which
reads: 
\begin{equation}
T_{\gamma }^{-}(\tau ,u)=T_{1/\gamma }(\tau ,u)=u^{1/\gamma -1}\exp (-\tau
u^{1/\gamma }),
\end{equation}
i.e. belong to the same class of functions than $T$'s themselves, but with $%
\gamma ^{*}=1/\gamma >1$. Note that the transforms $T_{\gamma }$ and $%
T_{\gamma }^{-}=T_{1/\gamma }$, lowering and rising the order of the FFPE to
the same amount are the inverse of each other: the Laplace-transform of $%
T_{\gamma }T_{\gamma }^{-}$ is $\exp (-u\tau )$ so that $T_{\gamma
}T_{\gamma }^{-}$ corresponds to a $\delta (\tau -t)$. Moreover, we have to
note that the $\tau $-integral of $T(\tau ,t)$, $N(t)=\int_{0}^{\infty
}T(\tau ,t)d\tau $ being an inverse Laplace-transform of $%
N(u)=\int_{0}^{\infty }u^{\gamma -1}\exp (-\tau u^{\gamma })d\tau =u^{-1}$,
is equal to 1 both for TETs ($\gamma <1$) and inverse ($\gamma >1$)
transforms, so that both the subordination and the inverse transformation
keep the overall normalization of the possible PDFss as functions of
coordinates.

The $T^{-}$-functions are not PDFss of $\tau $ since they may take negative
or even complex values. Let us fix some value of $\tau $ and consider the
limiting value of the integral $I(\tau )=\int_{0}^{\infty }T(\tau ,t)dt$,
which can be expressed in terms of $T(\tau ,u)$: $I(\tau )=\lim_{u\to
0}\left( u^{\gamma -1}e^{-\tau u^{\gamma }}\right) $. For $\gamma <1$ the
corresponding integral diverges being positive. On the other hand, for $%
\gamma >1$ $I(\tau )=0$, which means that the function $T(\tau ,t)$ either
changes its sign or vanishes identically. The last is not the case since the
integral $I_{1}(\tau )=\int_{0}^{\infty }tT(\tau ,t)dt=-\frac{d}{du}\left.
\left( u^{\gamma -1}e^{-\tau u^{\gamma }}\right) \right| _{u\rightarrow 0}$
still diverges for $1<\gamma <2$ (for larger values of $\gamma $ the
integrals $I_{n}(\tau )=\int_{0}^{\infty }t^{n}T(\tau ,t)dt$ with $n>\gamma
-1$ still diverge).

\section{The superdiffusive case}

Our derivation of the FFPE and its formal solution through subordination are
valid independently on the particular value of $\gamma $. The fact that $%
T_{\gamma }(\tau ,t)$ for $\gamma >1$ is not nonnegative does not mean that
the integral Eq.(\ref{Subord}) takes negative values: it solely means that
the non-negativity of the physical solutions of FFPEs does not follow from
the non-negativity of the physical solutions of the Fokker-Planck equation,
and that the variable $\tau $ can be no more interpreted as an internal time
governing the system's evolution. On the other hand, Eqs.(\ref{Pofu}) and (%
\ref{Lapla}) are still valid as a representation of a formal solution of the
FFPE. We shall refer to such formal solution as following from a
pseudo-subordination. In some special cases of pseudo-subordination one can
still can guarantee that the corresponding solution is a probability
distribution, as it is e.g. the case for force-free transport for $\gamma
\leq 2$, in other cases the solutions are not PDFs as it is e.g. for $\gamma
>2$.

\subsection{Pure superdiffusion: Relation to a wave equation}

Let us consider a purely diffusive situation without external force, 
\begin{equation}
F(x,t)=\frac{1}{2\sqrt{\pi Dt}}\exp \left( -\frac{x^{2}}{4Dt}\right) .
\end{equation}
The Laplace-transform of this function in $t$-variable reads: 
\begin{equation}
F(x,u)=\frac{1}{2}u^{-1/2}\exp \left( -\left| x\right| \sqrt{u}\right) .
\label{Difflap}
\end{equation}
Let us now use Eq.(\ref{Pofu}), and get $P(x,u)$ for arbitrary $\gamma $: 
\begin{equation}
P(x,u)=u^{\gamma -1}\tilde{F}(x,u^{\gamma })=\frac{1}{2}u^{\gamma /2-1}\exp
\left( -\left| x\right| u^{\gamma /2}\right) .
\end{equation}
The function $P(x,u)$ belongs to the class of functions $T(\tau ,t)$ given
by Eq.(\ref{Levy}), but with change of $\gamma $ to $\gamma /2$: 
\begin{equation}
P(x,t)=\frac{1}{2}T_{\gamma /2}(\left| x\right| ,t).  \label{Superdiff}
\end{equation}
Note that Eq.(\ref{Superdiff}) gives the representation of the
superdiffusive propagators in terms of the L\'{e}vy-functions, which
simplifies the general result of Ref.\cite{MeK1}. Since we know that $%
T_{\gamma }(\xi ,t)$ is a positive function of its both variables for $t>0$
and $\gamma \leq 1$, in the case of free propagation $P(x,t)$ is positive
for all $\gamma <2.$ The case $\gamma =2$ corresponding to 
\begin{equation}
P(x,t)=\frac{1}{2}\delta (\left| x\right| -t)  \label{delta}
\end{equation}
describes a special case of ballistic propagation. The equations of index $%
\gamma >2$ (describing a process which is {\it faster} than ballistic one)
do not correspond to transport of positive probabilities, since the
functions $T_{\gamma }$ with $\gamma >2$ are no more non-negative.

We have seen that although the non-negativity of the solution is not
mathematically guaranteed by the FFPE with $\gamma >1$ itself, the equation
still can possess physically reasonable positive solutions describing
superdiffusive transport. Let us discuss now the reasons, why it is so. Let
us note that the solution $P(x,u)=u^{-1/2}\exp \left( -\left| x\right| \sqrt{%
u}\right) $ itself can be considered as subordinated to a process described
by Eq.(\ref{delta}) (corresponding to $\Psi (x,u)=\sqrt{\pi }\exp \left(
-\left| x\right| u\right) $ ) with a ''subdiffusive'' subordination function 
$T_{1/2}(\tau ,t)$, so that the whole process can be considered as a
superposition of two subordination transformations, leading to the overall
behavior with $\gamma ^{*}=\gamma /2$. The process subordinated to a $\delta 
$-functional form under operational time given by $T_{\gamma /2}(\tau ,t)$
is, of course, exactly the solution given by Eq.(\ref{Superdiff}) discussed
before.

We note here that the two $\delta $-pulses described by Eq.(\ref{delta}) are
a solution of a wave equation (WE), 
\begin{equation}
\frac{\partial ^{2}\Psi }{\partial t^{2}}=\frac{\partial ^{2}\Psi }{\partial
x^{2}}.
\end{equation}
The solution Eq.(\ref{delta}) is not a Green's function solution of a wave
equation (known to be $G(x,t)=\frac{1}{2}\theta (t-\left| x\right| )$ in one
dimension, see Ref.\cite{MoF}) but a solution corresponding to a different
initial condition, namely to $G(x,t)\rightarrow \delta (x)\delta ^{\prime
}(t)$. The reason for this is easy to understand: The limiting equation for
the Green's function of a FFPE with $\gamma \rightarrow 2$ is not a
wave-equation, but a first-order integro-differential form, 
\begin{equation}
\frac{\partial F}{\partial t}=\int_{0}^{t}\frac{\partial ^{2}F}{\partial
x^{2}}dt^{\prime }+\delta (x)\delta (t),  \label{Idi}
\end{equation}
which is obtained from a wave equation by temporal integration.

\subsection{A problem of superdiffusion with drift}

Let us now consider processes being pseudo-subordinated to diffusive motion
under time-independent homogeneous external force (i.e. the solutions of
FFPEs 
\begin{equation}
\frac{\partial P}{\partial t}=~_{0}D_{i}^{\alpha }\left[ \mu f\frac{\partial 
}{\partial x}P(x,t)+D\frac{\partial ^{2}}{\partial x^{2}}P(x,t)\right] 
\end{equation}
with $\alpha =1-\gamma <0$). The Laplace-transform of the corresponding
Green's function of FPE, 
\begin{equation}
F(x,t)=\frac{1}{2\sqrt{\pi Dt}}\exp \left[ -\frac{\left( x-\mu ft\right) ^{2}%
}{4Dt}\right] ,
\end{equation}
reads: 
\begin{eqnarray}
&&F(x,u)=  \nonumber \\
&=&\frac{\exp (\frac{\mu fx}{2D})}{2\sqrt{\pi D}}\int_{0}^{\infty }\exp
\left[ -\left( \frac{\mu ^{2}f^{2}}{4D}+u\right) t-\frac{x^{2}}{4D}%
t^{-1}\right] \frac{dt}{\sqrt{t}}  \nonumber \\
&=&\frac{\exp (2\zeta \lambda )}{2\sqrt{D}}\frac{1}{\sqrt{\zeta ^{2}+u}}\exp
\left[ -\sqrt{(\zeta ^{2}+u)\lambda ^{2}}\right]   \label{Moving}
\end{eqnarray}
where the variables $\lambda =x/2\sqrt{D}$ and $\zeta =\mu f/2\sqrt{D}$ ($%
\zeta >0)$ are introduced (see 2.3.16.2 of Ref.\cite{BP}). Applying the $%
T_{2}$-transformation to Eq.(\ref{Moving}) we get: 
\begin{equation}
P_{2}(x,u)=\frac{\exp (2\zeta \lambda )}{2\sqrt{D}}\frac{u}{\sqrt{\zeta
^{2}+u^{2}}}\exp \left[ -\sqrt{(\zeta ^{2}+u^{2})\lambda ^{2}}\right] .
\label{P2}
\end{equation}
Let us show that $P_{2}(x,u)$ is not a Laplace-transform of a probability
distribution. Note that a Laplace-transform $f(u)$ of any nonnegative
function $f(t)$ must be an absolutely monotone function, i.e. $(-1)^{n}\frac{%
d^{n}}{du^{n}}f(u)\geq 0$ must hold for all $u$ and $n$. To prove this it is
easy to see that 
\begin{equation}
\frac{d^{n}}{du^{n}}f(u)=\int_{0}^{\infty
}f(t)e^{-ut}dt=(-1)^{n}\int_{0}^{\infty }t^{n}f(t)e^{-ut}dt.
\end{equation}
On the other hand, the first $u$-derivative of $P_{2}(x,u)$, changes its
sign at $u$ being a root of the equation 
\begin{equation}
\zeta ^{2}\sqrt{(u^{2}+\zeta ^{2})\lambda ^{2}}-u^{2}(u^{2}+\zeta ^{2})=0.
\end{equation}
The existence of positive roots of this equation for any $\zeta \neq 0$ is
clear since for $u$ small the overall expression (whose sign is the same as
the sign of $\frac{d}{du}P_{2}(x,u)$) is positive, and for large $u$ it is
negative. Note that the function gets to be absolutely monotone only when $%
\zeta =0$, i.e. only in the case of free diffusion. This observation is of
extreme importance since it shows that while the TETs ($T_{\gamma }\,$with $%
\gamma <1$), lowering the order of FFPE, always lead to reasonable physical
solutions, the inverse transformations, rising the order of FFPE do not
always do so.

Note that all functions $P_{\gamma }(x,t)$ obtained from diffusion with
drift under pseudo-subordination are not probability distributions for all $%
\gamma >1$. The Laplace-transform of the corresponding functions read 
\begin{equation}
P_{\gamma }(x,u)=\frac{\exp (2\zeta \lambda )}{2\sqrt{D}}\frac{u^{\gamma -1}%
}{\sqrt{\zeta ^{2}+u^{\gamma }}}\exp \left[ -\sqrt{(\zeta ^{2}+u^{\gamma
})\lambda ^{2}}\right] .
\end{equation}
The first derivative of $P_{\gamma }(x,u)$ changes its sign at $u$ being a
positive root of 
\begin{eqnarray}
&&((\gamma -2)u^{\gamma }+(2\gamma -2)\zeta ^{2})\sqrt{(u^{\gamma }+\zeta
^{2})\lambda ^{2}}  \nonumber \\
- &&\gamma u^{\gamma }(u^{\gamma }+\zeta ^{2})=0,
\end{eqnarray}
which function is positive for small $u$ and negative for larger ones. Thus,
the solutions of the FFPE of the type of a Fokker-Planck equation with $%
\gamma >1$ under homogeneous, constant force are not probability
distributions. Similar conclusions were drawn when considering the
particle's motion in a harmonic potential \cite{Metz}.

\section{Superdiffusive case: Subordination to a generic transport equation}

In Sec. 5.1 we have seen that the solutions of a diffusion equation
(representing a behavior of a stochastic process) are subordinated to
deterministic dynamics, described by a simple propagation of pulses with
constant velocity and given by a wave equation. Is the wave equation (i.e.
limiting superdiffusive FFPE with $\gamma =2$ for a force-free situation $%
f=0 $) a very special case, or there are some forms with $f\neq 0$ which
still lead to reasonable solutions?

It is clear that the second-order partial differential equation to whose
solutions the solution of FFPEs could be subordinated would read: 
\begin{equation}
\frac{\partial ^{2}P}{\partial t^{2}}=-\frac{\partial }{\partial x}\left[
A(x)P\right] +\frac{\partial ^{2}}{\partial x^{2}}\left[ B(x)P\right] .
\label{NPCE}
\end{equation}
Eq.(\ref{NPCE}) includes the wave equation as a special case. Eq.(\ref{NPCE}%
) will be called the generic transport equation (GTE), and, parallel to a
wave equation, has a dynamical (deterministic) nature. This equation (being
a close relative of Liouville equation) was considered by the author in a
different context in Ref.\cite{Sokolov}: The GTE appears when restoring
temporal dependence in a Pope-Ching equation for stationary random
processes, Ref.\cite{PiC}. The meaning of prefactors here is: $%
A(x)=\left\langle \ddot{x}(x)\right\rangle $ and $B(x)=\left\langle \dot{x}%
^{2}(x)\right\rangle $, so that for a physical particle they are
proportional to the acting force and to the particle's mean kinetic energy.

Let us remind the procedure of derivation of GTE given in Ref. \cite{Sokolov}%
. The PDF of $x$, $p_{x}(x)$, is obtained as an ensemble-average (e.g. over
the initial conditions) of the realizations for each of which

\begin{equation}
p(x,t)=\delta (X(t)-x)  \label{Eq.2}
\end{equation}
where $X(t)$ represents the law of motion. The coarse-grained probability is
then given by $p(x)=\left\langle p(x,t)\right\rangle $. Differentiating Eq.(%
\ref{Eq.2}) with respect to time one gets

\begin{equation}
\frac{\partial p}{dt}=-\dot{X}\frac{\partial p}{\partial x}=-\frac{\partial 
}{\partial x}\left( \dot{X}p\right) .  \label{Eq.3}
\end{equation}
since $X$ is independent on $x$. Note that Eq.(\ref{Eq.3}) is a Liouville
equation, and the derivation here is parallel to one given in Ref.\cite
{Gardiner}. Applying the same procedure for the second time we get:

\begin{eqnarray}
\frac{\partial ^{2}p}{\partial t^{2}} &=&-\frac{\partial }{\partial x}\frac{%
\partial }{\partial t}\left( \dot{X}p\right) =\frac{\partial }{\partial x}%
\left( \ddot{X}p\right) +\frac{\partial }{\partial x}\left[ \dot{X}\frac{%
\partial }{\partial x}\left( \dot{X}p\right) \right]   \nonumber \\
&=&-\frac{\partial }{\partial x}\left( \ddot{X}p\right) +\frac{\partial ^{2}%
}{\partial x^{2}}\left( \dot{X}^{2}p\right) ,  \label{Eq.4}
\end{eqnarray}
which equation is, of course, the as exact as the Liouville one. The GTE
follows after the ensemble averaging, under which the corresponding
conditional means appear instead of the instantaneous velocity and
acceleration, so that Eq.(\ref{Eq.4}) reduces to Eq.(\ref{NPCE}). Note that
the GTE can be useless but is never false: its solutions describe all
possible motions and are both dynamically and thermodynamically sound. These
solutions are probability densities. On the other hand, the prefactors $A$
and $B$ arise as (nonequilibrium) ensemble averages and depend on what
ensemble is used and thus on the initial conditions: a simple example of
this fact is considered below. The absence of the physical solution of Eq.(%
\ref{NPCE}) means that the corresponding thermodynamical forces and kinetic
coefficients defining $A(x)$ and $B(x)$ are incompatible with each other or
with the initial conditions and would never appear as thermodynamical means.
Moreover, even if the system as a whole is homogeneous and its physical
properties are time-independent, the coefficients $A(x,t)$ and $B(x,t)$ can
be time-dependent and will relax to the equilibrium values not faster than
the distribution itself relaxes to its equilibrium form, which explains the
unphysical sort-time behavior of the solutions of superdiffusive FFPE in
harmonic potential found in Ref. \cite{Metz}.

As an example of a processes subordinated to a solution of GTE let us
consider a simple oscillatory process taking place in the operational time
of the system. The dynamic equation of the oscillator is 
\begin{equation}
\ddot{x}=-\omega x.
\end{equation}
Let us consider the situation when the oscillator starts with zero velocity
at $x=-a$ so that $A(x)=-\omega ^{2}x$, and $B(x)=\omega ^{2}(a^{2}-x^{2})$.
Our process is described in operational time by a GTE

\begin{equation}
\frac{\partial ^{2}F}{\partial \tau ^{2}}=\frac{\partial }{\partial x}\left(
\omega ^{2}xF\right) +\frac{\partial ^{2}}{\partial x^{2}}\left[ \omega
^{2}(a^{2}-x^{2})F\right]   \label{Dynamic}
\end{equation}
with the initial conditions $F(x,0)=\delta (x+a)$, and $\left. \frac{%
\partial F(x,\tau )}{\partial \tau }\right| _{\tau =0}=0$, whose solution,
as anticipated, reads $F(x,\tau )=\delta (x+a\cos \omega \tau )$. Note that
the coefficient $B$ depends explicitly on $a$, so that the form of equation
depends on the initial energy of the oscillator. Equation (\ref{Dynamic}) is
incompatible with any combination of initial conditions not leading to the
same amplitude of oscillations, i.e. whenever one supposes $\omega
^{2}x^{2}(0)+\dot{x}^{2}(0)\neq a^{2}$, and would lead in this case to
negative or complex PDFs. The solution subordinated to $F(x,\tau )$ reads: 
\begin{eqnarray}
P(x,u) &=&\int_{0}^{\infty }\delta (x+a\cos \omega \tau )u^{\gamma -1}\exp
(-\tau u^{\gamma })d\tau   \nonumber \\
&=&\frac{u^{\gamma -1}}{\omega a\sqrt{1-x^{2}/a^{2}}}\exp \left[ \frac{1}{%
\omega }\arccos \left( -\frac{x}{a}\right) u^{\gamma }\right] \times  
\nonumber \\
&&\times \sum_{n=0}^{\infty }\exp (-\frac{\pi }{\omega }nu^{\gamma }) \\
&=&\frac{u^{\gamma -1}}{\omega a\sqrt{1-x^{2}/a^{2}}}\frac{\exp \left[ \frac{%
1}{\omega }\arccos \left( -\frac{x}{a}\right) u^{\gamma }\right] }{1-\exp (-%
\frac{\pi }{\omega }u^{\gamma })}.  \nonumber
\end{eqnarray}
For $t\rightarrow \infty $ (full dephasing) the corresponding PDF tends to a
PDF to find an oscillating point at coordinate $x$. Thus, for $u\rightarrow 0
$%
\begin{equation}
P(x,u\rightarrow 0)\simeq \frac{1}{u}\frac{1}{\pi a\sqrt{1-x^{2}/a^{2}}}
\end{equation}
which corresponds to 
\begin{equation}
P(x,t\rightarrow \infty )\simeq \frac{1}{\pi a\sqrt{1-x^{2}/a^{2}}},
\label{As1}
\end{equation}
a well-known solution for the invariant PDF for a classical harmonic
oscillator. Thus, the equations subordinated to GTE may describe partly
coherent phenomena: their physical relation to a wave equation gets evident.
It is interesting to mention that the Fokker-Planck equation with the same
coefficients as Eq.(\ref{Dynamic}) describes a harmonic oscillator in which
the diffusion takes place in an inhomogeneous temperature field, $T(x)\simeq 
\sqrt{1-x^{2}/a^{2}}$, and such a diffusive solution stays physically sound
both under subordination and under pseudo-subordination up to $\gamma =2$.

The discussion above shows that for all $A(x)$ and $B(x)$ and initial
conditions for which Eq.(\ref{NPCE}) has real, nonnegative, normalizable
solutions, the free relaxation of a complex system under FFPE dynamics can
be described as its deterministic development in its own (operational) time.
The inverse (as we have proved on an example of a diffusion with drift) is
not the case. Our considerations leave open the question whether all
physically sound solutions of superdiffusive FFPEs are subordinated to ones
of GTE, or this class is wider and includes some functions which are
probability densities for $1<\gamma <\gamma ^{*}$ and cease to be
probability densities for $\gamma ^{*}<\gamma <2$.

It is important to stress that the fact that the solutions of superdiffusive
FFPEs in general are not probability densities, does not devaluate the FFPEs
as an instrument for description of complex relaxation phenomena, but shows
that many combinations of thermodynamical forces, kinetic coefficients and
memory functions will never appear as thermodynamic ensemble means. This
means that the forces and kinetic coefficients can not be invented ad hoc,
but must follow either from experiments or from microscopic considerations.
In the case when the correct thermodynamical forces and the impedance of the
system are known, its FFPE is uniquely determined.

The consideration of GTE explains also our finding that the solutions for
the force-free transport with $\gamma >2$ are not non-negative. Such
equations would describe processes subordinated to the solutions of the
exact third-order transport equation. The exact equation with $\gamma =3$
arising from applying a Liouville operator $\frac{\partial }{\partial t}+%
\dot{x}\frac{\partial }{\partial t}$ to the PDF $P$ three times is a
trinomial construct, with correlated coefficients in front of the first, the
second, and the third spatial derivatives. This third-order equation is a
generic form for transport equations of higher order. The equations
subordinated to this one will have third-order structure in spatial
variables and will be hardly a helpful tool, since they do not have any
known classical counterpart whose solutions may be used for building new
ones.

\section{Conclusions}

Fractional Fokker-Planck equations (FFPE) with additional fractional time
derivative in front of a normal Fokker-Planck operator appear within a usual
linear-response theory when describing systems showing strange kinetics. We
show that such form of FFPEs describes systems in a contact with a heat
bath, since the noise in such systems in equilibrium (for $t\rightarrow
\infty $) fulfills the Nyquist theorem. Many other forms (e.g. with temporal
fractional derivatives of different orders in front of first- and
second-order spatial derivatives) are thus ruled out as appropriate for
describing situations close to equilibrium, although they may be appropriate
for many other transport processes, as e.g. dispersion by flows. Using the
fact that the solutions of subdiffusive FFPEs with time-independent
coefficients are subordinated to those of normal Fokker-Planck equations, we
show that the FFPE solutions are probability densities in all cases when the
usual Fokker-Planck equation has physical solutions. Thus, the free
relaxation of a complex system under FFPE dynamics can be described as its
development in its own (operational) time. The superdiffusive FFPEs do not
posses physical solutions for arbitrary choice of force and diffusion
coefficient. This does not devaluate the FFPEs as a tool for description of
superdiffusive processes, but stresses the fact that the corresponding
combination of thermodynamical forces and memory functions can never emerge
as a thermodynamical ensemble average: the corresponding phenomenological
equations have to be handled with care.

\section{Acknowledgments}

The author is indebted to Mr. M. Kassmann, Prof. A. Blumen and Prof. J.
Klafter for fruitful discussions, to Dr. R. Metzler for communicating his
recent results, and to Mr. J. Stampe for careful reading of the manuscript.
Financial support by GIF, by the Deutsche Forschungsgemeinschaft through the
SFB 428 and by the Fonds der Chemischen Industrie is gratefully acknowledged.

\end{document}